# OVERDOPED CUPRATES WITH HIGH TEMPERATURE SUPERCONDUCTING TRANSITIONS

M. Marezio,[1] O. Chmaissem,[2,3] C. Bougerol,[4] M. Karppinen,[5] H. Yamauchi,[5] T.H Geballe[6],*

[1]*CRETA/CNRS, Avenue des Martyrs, Grenoble 38042, France*
[2]*Physics Department, Northern Illinois University, DeKalb, IL 60115, USA*
[3]*Materials Science Division, Argonne National Laboratory, Argonne, IL 60439, USA*
[4]*CEA-CNRS Nanophysics and semiconductors group, Institut Néel, CNRS, 25 Avenue des Martyrs, 38042 Grenoble, Cedex 9, France*
[5]*Department of Chemistry, Aalto University, FI-00076 Aalto, Finland*
[6]*Department of Applied Physics and Materials Science, Stanford University, Stanford, California 94305, USA*

Abstract

Evidence for High $T_c$ cuprate superconductivity is found in a region of the phase diagram where non-superconducting Fermi liquid metals are expected. Cu valences estimated independently from both x-ray absorption near-edge structure (XANES) and bond valence sum (BVS) measurements are > 2.3 and are in close agreement with each other for structures in the homologous series $(Cu_{0.75}Mo_{0.25})Sr_2(Y,Ce)_sCu_2O_{5+2s+\delta}$ with s = 1, 2, 3, and 4. The s = 1 member, $(Cu_{0.75}Mo_{0.25})Sr_2YCu_2O_{7+\delta}$, $0 \leq \delta \leq 0.5$, is structurally related to $YBa_2Cu_3O_7$ in which 25% of the basal Cu cations [i.e. those in the chain layer] are replaced by Mo, and the Ba cations are replaced by Sr. After oxidation under high pressure the s = 1 member becomes superconducting with $T_c$ = 88K. The Cu valence is estimated to be ~2.5, well beyond the ~2.3 value for which other High-$T_c$ cuprates are considered to be overdoped Fermi liquids. The increase in valence is attributed to the additional 0.5 oxygen ions added per chain upon oxidation. The record short apical oxygen distance, at odds with current theory, suggests the possibility of a new pairing mechanism but further experiments are urgently needed to obtain more direct evidence. From the structural point of view the members with s ≥ 2 are considered to be equivalent to single-layer cuprates because the $T_c$ is independent of the thickness of the insulating fluorite-like blocks on going from the s = 2 to the s = 5. All have $T_c$ ~ 56 K which is significantly higher than expected because they also have higher than expected Cu valences. The XANES-determined valences normalized to give values in the CuO2 layers are 2.24, 2.25, and 2.26 for s = 2, 3, and 4, while the BVS values determined for the valence in the $CuO_2$ layer alone are 2.31-2.34 for the s = 2 and 3 members. No evidence for periodic ordering has been detected by electron diffraction and high resolution imaging studies. The possibility that the charge reservoir layers are able to screen long range coulomb interactions and thus enhance Tc is discussed.

*Corresponding author: T.H.Geballe: Geballe@stanford.edu

**Introduction**:

During the past eight years the first six members of the homologous series $(Cu_{0.75}Mo_{0.25})Sr_2(Ce,Y)_sCu_2O_{5+2s+\delta}$ have been successfully synthesized and investigated.[1-6] The reported results indicate that the Cu valence is anomalously higher than in the well-known high-$T_c$ cuprates such as $YBa_2Cu_3O_7$. The s = 1 member, $(Cu_{0.75}Mo_{0.25})Sr_2YCu_2O_{7+\delta}$, is structurally the same as $YBa_2Cu_3O_7$ but where 25% of the basal Cu cations [i.e. those in the chain layer] are replaced by Mo, and the Ba cations are replaced by Sr. The formulae for the members corresponding to s = 2, 3, 4, 5, 6 are: $(Cu_{0.75}Mo_{0.25})Sr_2(Ce,Y)_2Cu_2O_{9+\delta}$, $(Cu_{0.75}Mo_{0.25})Sr_2(Ce,Y)_3Cu_2O_{11+\delta}$, $(Cu_{0.75}Mo_{0.25})Sr_2(Ce,Y)_4Cu_2O_{13+\delta}$, $(Cu_{0.75}Mo_{0.25})Sr_2(Ce,Y)_5Cu_2O_{15+\delta}$, and $(Cu_{0.75}Mo_{0.25})Sr_2(Ce,Y)_6Cu_2O_{17+\delta}$, respectively. There is a substantial difference between the familiar Hg-m2(n-1)n and Tl-m2(n-1)n homologous series and the present $(Cu_{0.75}Mo_{0.25})$-12s2 series. In the Hg- and Tl-series n represents the number of $(Ca)(CuO_2)$ blocks in the unit cell, i.e. n = 1, 2, 3, 4 means 1, 2, 3, 4 $CuO_2$ layers per unit cell. In the $(Cu_{0.75}Mo_{0.25})$-12s2 series there are only 2 $CuO_2$ layers per unit cell; s =1 stands for a single layer of Y, and s > 1 stands for s-1 additional insulating $(Ce,Y)(O_2)$ layers that are in the $CaF_2$ fluorite configuration.

The increase in $T_c$ on going from the n = 1 to n = 3 members of the Hg- or Tl-based series is believed to be due to interlayer pair tunneling between the $CuO_2$ layers that suppresses phase fluctuations and possibly, for n=3, because of the interactions between the underdoped inner layer and the overdoped outer layers can enhance Tc. In contrast, the decrease from 88 to an almost constant 56 K observed on going from the s = 1 to s > 1 for members of the present homologous series is due to the thickening of the insulating fluorite-like layers that prevents interlayer tunneling. The almost constant $T_c$ for s >1 can be understood because the coupling between unit cells is achieved by the same reservoir $(SrO)(Cu_{0.75}Mo_{0.25}O_{1+\delta})(SrO)$ block present in every member of the series. Thus all the members of this homologous series with s >1 are comparable to single layer 214 cuprates; to a first approximation they can be considered as single-layer cuprates with $T_c$s ≈ 56 K that are significantly higher than optimally-doped single-layer cuprates with the exception of $Sr_2CuO_{4-x}$[8-11] and $HgBa_2CuO_{4+\delta}$.[12]

**Previous Results:**

Powder x-ray diffraction and high resolution transmission electron microscopy show that the s = 2 member contains the following layered sequence:

$([Cu_{0.75}Mo_{0.25}]O_{1+\delta})(SrO)(CuO_2)(Y,Ce)(O_2)(Y,Ce)(CuO_2)(SrO)([Cu_{0.75}Mo_{0.25}]O_{1+\delta})$.[2] The sequence of the higher members is obtained by inserting additional two-layer blocks of $(O_2)(Y,Ce)$ between the $(CuO_2)$ layers (see Fig. 1). The as-synthesized (AS) samples were not superconducting. They became so by means of high-pressure oxygenation (HPO) carried out at 5 GPa and 500 °C in the presence of $KClO_3$ acting as the oxidant. With increasing ratio of $KClO_3$ to the cuprate phase the c axis gradually decreased and $T_c$ increased up to 88 K for the s = 1, and to ~ 56 K for all the higher members. The thickness of the inserted fluorite block increases from 6.4 Å for s = 2 to 11.4 Å for s = 4. The lattice parameters of the more recent synthesis of the s = 5 and 6 members[6] are in good agreement with those expected from the lower homologues although the slower reaction rates made further experiments problematic as the samples obtained were not completely single phase. Nevertheless, the s = 5 homologue became superconducting with a $T_c$ of ~55 K after a routine HPO treatment that is similar to the s < 5 samples. Field-cooled and zero field-cooled susceptibility measurements[2] give evidence for bulk superconductivity (volume fraction > 30%). Furthermore the lack of an appreciable Curie tail above Tc is evidence that there is little no magnetic second phase present. The Cu valence has been investigated by two independent methods, namely x-ray absorption near-edge spectroscopy (XANES) and the empirical bond-valence-sum (BVS) method that uses the bond distances determined from the neutron powder diffraction refinements. The BVS determined values for the Cu valence in the $CuO_2$ layer in the HPO samples are 2.45, 2.31, and 2.34 for the s = 1, 2, and 3, respectively. The XANES measured valence is an average of the Cu cations in the reservoir block and those in the $CuO_2$ layers. A procedure that relies on the resolved oxygen K-edge peaks in the chains and planes was used to obtain Cu valence in the $CuO_2$ layers alone[2]. The XANES valences for the HPO samples are 2.46, 2.24, 2.28, and 2.26 for s = 1, 2, 3, and 4. The Cu valences of the corresponding non-superconducting AS samples were found to be 2.16, 2.13, 2.14 and 2.14 for the s = 1, 2, 3 and 4, samples, respectively. The BVS valence gains credibility because of earlier work in $YBa_2Cu_3O_7$ where the BVS valence is in reasonable agreement with the values found later by neutron powder diffraction and thermoelectric power measurements.[13,14] It is plausible to assume that the BVS valence will also be a reliable estimate of the mobile charge density in the present series. The possibility that superconductivity occurs in a region of the phase diagram where it would be expected to be a non-superconducting normal Fermi liquid has been previously postulated to explain the properties of $Sr_2CuO_{4-x}$.[10,11] In 1991 Li Rukang et al. investigated a long list of possible fluorite-inserted cuprates. The formula of their compounds was $MSr_2(Ln,R)_2Cu_2O_y$ with M = Ti, V, Nb, Ta, Mo, W, Sn, Sb which corresponds to the s = 2 members of the respective homologous series. No superconductivity was found in any of these cuprates.[15,16] The success for the present series seems to be due to the replacement of 25% of Cu in the reservoir layer with Mo and to the extra oxygen that this substitution induces in the same layer.

When the Mo content was varied from 5% to 35% for the s = 3 member it was found that 25% represented the maximum concentration that can be incorporated.[3] At higher concentrations the c parameter remained constant and impurities began to appear in the powder pattern. A continuous increase of $T_c$ and the c parameter was observed from 5% to 25%. The same authors tried other cations such as Re, W, and Pb to partially substitute for the Cu of the reservoir layer. Only the first two substitutions yielded superconducting samples after high-pressure and high-temperature oxygenation. The (Cu,Re)-1232 compound showed a $T_c \approx 55$ K while the W counterpart showed a somewhat lower $T_c$ at 53 K. It is possible that W, being larger than Cu or Mo resulted in an ordered layer causing the decrease in $T_c$. The Pb-substituted 1232 sample did not show any superconductivity after the same oxygenation treatment. All the experimental evidences suggest that a high valence cation is needed in order to induce superconductivity in the oxygenated compounds.

**Results and Discussion:**

The density of mobile carriers generally accepted to be given by how much the formal valence of the Cu ions in the $CuO_2$ layers exceeds 2, usually denoted by δ. Attributing these holes to the Cu ions is just a simple way of keeping track of the charge density. $T_c$ is believed to follow a "universal" curve or "dome" which reaches a maximum at "optimal doping" that occurs for a valence of ~2.16 and returns to 0 as the valence approaches 2.28.[14] The present system shows that this behavior is likely not universal. The BVS and XANES analyses find the Cu valence to be ~2.45 for the s = 1 homologue that has $T_c$ = 88 K. The Cu valences of the s = 2, 3, and 4 members (with $T_c$s ≈ 56 K) are in the nominally overdoped region where the universal curve is, or is rapidly, approaching 0.

We interpret the above results to be evidence that the oxygenated charge reservoir blocks play an important role in defining the superconducting properties. The s = 1 member, as pointed out in the introduction, differs significantly from the well known $YBa_2Cu_3O_7$ structure by having 25 % of Cu cations in the basal chain layer replaced by octahedrally-coordinated $Mo^{6+}$ cations. Consequently, the basal layer of the Mo-substituted compounds contains more oxygen than those of the corresponding layer in $YBa_2Cu_3O_7$ or $YSr_2Cu_3O_7$. Room temperature structural refinements based on powder neutron diffraction data obtained at Argonne National Laboratory[17] show that the δ values for all members of the series vary from 0 for the AS samples to 0.5 for the HPO samples. The total oxygen content for the first member was found to be 7.36 for the non-superconducting sample and 7.56 for the sample with $T_c$ = 88 K.

The lack of ordering of the Cu and Mo cations in the basal layer is revealed by the electron diffraction and high resolution electron images for the s = 1, 2 and 3 members shown in Fig. 2 for the AS samples. The disorder should also apply to the HPO samples because, while the annealing under pressure increases oxygen content and thus the doping, it is not expected to cause cation diffusion. In Fig. 2a the electron diffraction pattern along to the [010] axis is shown for the AS s = 1 sample. No superstructure spots or diffuse scattering are visible. The same is true for other reciprocal directions. In Fig. 2b a high-resolution image taken with a JEOl -4000EX microscope operated at 400 kV for the same sample is shown. It can be seen that the experimental image is in good agreement with the simulated structure shown in the inset obtained by using the crystal structure reported in reference 17. Fig. 3a shows an electron diffraction pattern for the AS s = 3 sample. Four crystals were investigated and all produced the same results. The electron diffraction pattern is along the [010] zone axis. Neither superstructure spots nor diffuse streaks are visible. In Fig. 3b a high resolution micrograph is shown with the simulated image displayed in the inset. It can be seen from Table 1 that for the s = 3 member the two types of blocks separated by the $CuO_2$ layers have about the same thickness, ~8.5 Å, which is nicely confirmed by the high resolution image. Our data confirm the results reported in reference 1 for the sample with s = 3.

The failure to observe any long-range order between the Mo and Cu cations in the basal layer by the diffraction experiments is a strong indication that the $Mo^{6+}$ cations are in octahedral coordination. In this environment $Mo^{6+}$ has an ionic radius of 0.59 Å that is very close to that of $Cu^{2+}$ in square coordination, 0.57 Å, while the ionic radius of $Mo^{6+}$ in tetrahedral coordination is estimated to be 0.41 Å.[18]

It remains a challenge to discover the mechanism responsible for superconductivity in the current system. Recently, it has been shown theoretically that the off-site repulsive interactions that reduce $T_c$ in d-wave superconductors can be screened and thus $T_c$ will be enhanced if they are capacitively coupled to either a high polarizable or a metallic charge-reservoir layer.[19] In the present case the high-frequency susceptibility of the disordered-chain layer may be effective in providing the screening. Experimental investigations such as NMR are needed to support this model.

Fig. 4 shows a possible structural arrangement of the basal plane for all AS samples. The yellow oxygen octahedra surround the Mo cation which is placed at 000 in the subcell or the hypothetical quadruple-ordered unit cell. The positions at 100, 010 and 110 (or ½00, 0½0 and ½½0 of the quadruple cell) are occupied by Cu cations. The first two have a square coordination while the one at 110 (½½0) has the dumbbell coordination. The chemical formula of the four reduced samples are: $(Cu_{0.75}Mo_{0.25})Sr_2YCu_2O_7$, $(Cu_{0.75}Mo_{0.25})Sr_2(Y,Ce)_2Cu_2O_9$, $(Cu_{0.75}Mo_{0.25})Sr_2(Y,Ce)_3Cu_2O_{11}$, and $(Cu_{0.75}Mo_{0.25})Sr2(Y,Ce)_4Cu_2O_{13}$.

Fig. 5 represents the basal plane for all HPO samples. The additional oxygen can occupy only one empty position of every other subcell. No more oxygen can be incorporated otherwise the resulting O-O distances would become too short. Consequently the maximum theoretical value for $\delta$ in the HPO samples is 0.5. The occupancy factors of the oxygen sites obtained from the NPD refinements for the insulating AS and superconducting HPO phases as stated above correspond to a total oxygen content of 7.36 and 7.56 for the $s=1$ member and 9.01-9.41, 11.16-11.72 for the $s = 2, 3$ members, respectively.[17] For the $s = 4$ member the structural refinement based on neutron data was carried out only for the AS phase. There was not enough left sample to collect data for the HPO phase refinement. The oxygen content of $s =4$ AS sample was found to be 12.97. We have not yet made a systematic investigation to determine if there is an optimum doping level at some intermediate value.

It is important to point out that the apical Cu-O distance in the $s = 1$ AS compound is 2.295(4) Å, very close to that found in superconducting $YBa_2Cu_3O_{6.95}$, 2.301 Å for which $T_c = 90$ K.[13] After the HPO treatment resulting in superconductivity ($T_c = 88$ K), the apical distance in $(Cu_{0.75}Mo_{0.75})$-1212 decreases to 2.15(2) Å, contrary to the two-orbital model of Sakakibara et al[20] or to any other theory that finds Tc increases with apical O distance.

According to Jorgensen et al.[21] the apical oxygen distance of 2.15 Å together with the O-Cu-O corrugation angle of 166° in the $CuO_2$-square layers suggest that the reservoir block of the $s = 1$ member, $(Cu_{0.75}Mo_{0.75})Sr_2YCu_2O_{7.5}$, has a metallic character. The idea of metallic reservoir blocks is not new. As early as 1996 Tallon and coworkers[22] inferred from muon spin relaxation studies that a 50-fold increase in irreversibility field observed when 25% of the Hg cations are substituted by Re and all the Ba cations are replaced by Sr in $HgBa_2CaCu_2O_{6+\delta}$ could be explained by the metallization of the $(Hg_{0.75}Re_{0.25})O_{4+\delta}$ reservoir layer.

To explain why $T_c$ decreases on going from the s = 1 to the s = 2 member (87 to 56 K) and then remains practically constant for the s = 3 and 4 we report in Table 2 the thickness of the two blocks existing in the structure of these compounds. These structures can be viewed as built of two types of blocks: one being the $CuO_2$(fluorite-block of different thickness)$CuO_2$, and one of about constant thickness: $CuO_2(SrO)(Cu_{0.75}Mo_{0.25})(SrO)CuO_2$. The fluorite block does not exist in the s = 1 member or in $YBa_2Cu_3O_{6.95}$. In both cases it consists of only one yttrium layer. Two observations can be made from Table 1: 1) the oxygen insertion increases the thickness of the $CuO_2$(fluorite-block)$CuO_2$ block even though the extra oxygen is not incorporated into that block. 2) The thickness of the reservoir block, which is the same for all members, decreases when the extra-oxygen is incorporated by the HPO treatment. These observations indicate that these compounds have a strong ionic character because by inserting negative oxygen ions the Coulomb interactions definitely prevail against the size effect. The same observation can be made for $YBa_2Cu_3O_{6+x}$ which is included in Table 2 for comparison. Moreover, the same structure of the "reservoir block" for all members with inserted-fluorite blocks can explain why $T_c$ remains constant for s = 2, 3, and 4 members.

**Conclusions and future work:**

The high valence value of 2.45 obtained for the s = 1 HPO sample indicates, as has already been suggested from earlier experiments in another system,[10,11] that superconductivity exists in the very overdoped region of the cuprate phase diagram. We suggest that $T_c$ can be enhanced because the longer range repulsive Coulomb interactions in the $CuO_2$ layer are screened by the highly polarizable charge reservoir layers. However since no measurements of the superfluid density have been made there remains the possibility that an undetected inhomogeneity such as stripe formation might explain the results without requiring new physics. It is an outstanding challenge for future work to make decisive measurements. Further doping is needed to establish whether or not a second dome exists and whether optimal doping has been achieved. The increase in $T_c$ found upon annealing which causes a marked decrease in the apical O distance while the Cu-O bond length in the plane remains almost constant cannot be explained by any current theory.[20]

Many other cuprates related to the present series remain to be investigated. The cation substitutions M = Ti, V, Nb, Ta, Mo, W, Sn, Sb reported by Li Rukang et al.[15,16] are all s = 2 members of the respective $MSr_2(Ln,R)_2Cu_2O_y$ homologous series. The Sr version of these superconducting cuprates is essential because Ba is too big to allow for the extra oxygen that is needed. In order to keep Cu in the reservoir block these substitutions should be only partial. The preliminary results on Re substitution[3] seem to suggest that highly oxidized cations are good candidates for producing enhanced superconductivity in the system $YSr_2Cu_3O_y$.

**Acknowledgments:**

Work at Argonne was supported by the U.S. Department of Energy, Office of Science, Materials Science and Engineering Division, under contract No. DE-AC02-06CH11357. The work at Aalto University was supported by Academy of Finland (Nos. 126528 and 255562) and Tekes (No. 1726/31/07). The work at Stanford was supported in part by the Airforce Office of Scientific Research (AFOSR) under grant FA9550-09-1-0583. THG would like to acknowledge helpful comments from Sri Raghu, Steve Kivelson, and Aharon Kapitulnik.

22. Tallon, JL, Bernhard, C., Niedermayer, Ch., Shimoyama, J., Hahakura, S., Yamaura, K., Hiroi, Z., Takano, M. & Kishio, K., A new approach to the design of high-T-c superconductors: Metallised interlayers. Journal of Low Temperature Physics **103**, 1379-1384 (1996)

**Figure Captions**

Fig. 1: Structure of (Cu,Mo)-1222. For clarity, only the oxygen atoms within the fluorite-like $CeO_2$ blocks are plotted.

Fig. 2: (a) Electron diffraction pattern corresponding to the basal plane of the AS sample $(Cu_{0.75}Mo_{0.25})$-1212. No extra spots are visible which would indicate the existence of a superstructure due to the Cu and Mo long-range ordering in the basal plane. (b) A high resolution image showing the s =1 layered structure in agreement with reference (14). A simulated structure is shown in the inset.

Fig. 3: (a) a [010] electron diffraction projection for the AS sample (Cu,Mo)-1232. No long range ordering observed between the Mo and Cu cations. (b) A high resolution image showing the s = 3 layered structure in agreement with reference (14). A simulated structure is shown in the inset.

Fig. 4: A possible structural arrangement of the basal plane $(Cu_{0.75}Mo_{0.75})O_{1+\delta}$, $\delta = 0$ for all the AS samples. The subcell is indicated. The extra oxygen atoms are disordered over the xy0 and the more symmetrical ½y0 positions. For only the AS s = 4 sample, the oxygen atoms are in the xy0 positions with a 0.125 occupancy factor. Spheres are: Mo (yellow), Cu (blue), and O at (red at ½y0 and green at xy0). Additional oxygen atoms (green) are shown at the corners of the Mo octahedral and on top of the Cu spheres.

Fig. 5: A possible structural arrangement of the basal plane $(Cu_{0.75}Mo_{0.75})O_{1+\delta}$, $\delta = 0.5$ for all the HPO samples. The subcell is indicated. The oxygen atoms are all in the ½y0 positions with a 0.375 occupancy factor. This occurs only in the HPO s = 1 sample. In all others members the oxygen is disordered over the ½y0 and the less symmetrical xy0 positions. Same color scheme as in Fig. 4.

**Table 1**

Thickness in Å of the 2 blocks separated by the $CuO_2$ layers, which make up the structure of the homologous series $(Cu_{0.75}Mo_{0.25})$-12s2

| Sample | $CuO_2$(fluorite-block)$CuO_2$ | $(CuO_2)(SrO)(Cu_{.75}Mo_{.25}O_x)(SrO)(CuO_2)$ |
|---|---|---|
| s=1 AS | 3.308* | 8.242 |
| s=1 HPO | 3.434* | 8.037 |
| s=2 AS | 5.998 | 8.283 |
| s=2; HPO | 6.064 | 8.161 |
| s=3; AS | 8.746 | 8.237 |
| s=3; HPO | 8.801 | 8.104 |
| s = 4; AS | 11.441 | 8.231 |
| $YBa_2Cu_3O_6$ | 3.304* | 8.513* |
| $YBa_2Cu_3O_{6.95}$ | 3.391* | 8.269* |

* In $YBa_2Cu_3O_6$ and $YBa_2Cu_3O_{6.95}$ the fluorite blocks do not exist, the two $CuO_2$ layers are separated by a single Y layer. Furthermore, the basal layer contains only Cu cations.

**Table 2**

The Cu2 cations bond-valence-sums (BVS) calculated from the structural refinements based on neutron diffraction powder data[17] compared with the valence obtained from XANES data.[2] The s values (1, 2, 3, 4) indicate the four members of the homologous series in the as-prepared (AS) and HPO state.

|  | s = 1 | | s = 2 | | s = 3 | | s = 4 | |
| --- | --- | --- | --- | --- | --- | --- | --- | --- |
|  | AS | HPO | AS | HPO | AS | HPO | AS | HPO |
| v(Cu2) from BVS | 2.21 | 2.45 | 2.22 | 2.31 | 2.22 | 2.34 | 2.20 | |
| v(Cu) from XANES | 2.16* | 2.46 | 2.13 | 2.24 | 2.14 | 2.25 | 2.14 | 2.26 |
| $\rho(CuO_2)$ | | 0.53 | | 0.28 | | 0.26 | | 0.27 |
| $T_C$ | | 87 K | | 56 K | | 54.5 K | | 55.5 K |

*As stated in reference 5 the oxygen content for this sample was determined by iodometric titration and exactly the same value of 2.16 was obtained. This is an indirect proof that the samples are of high quality.

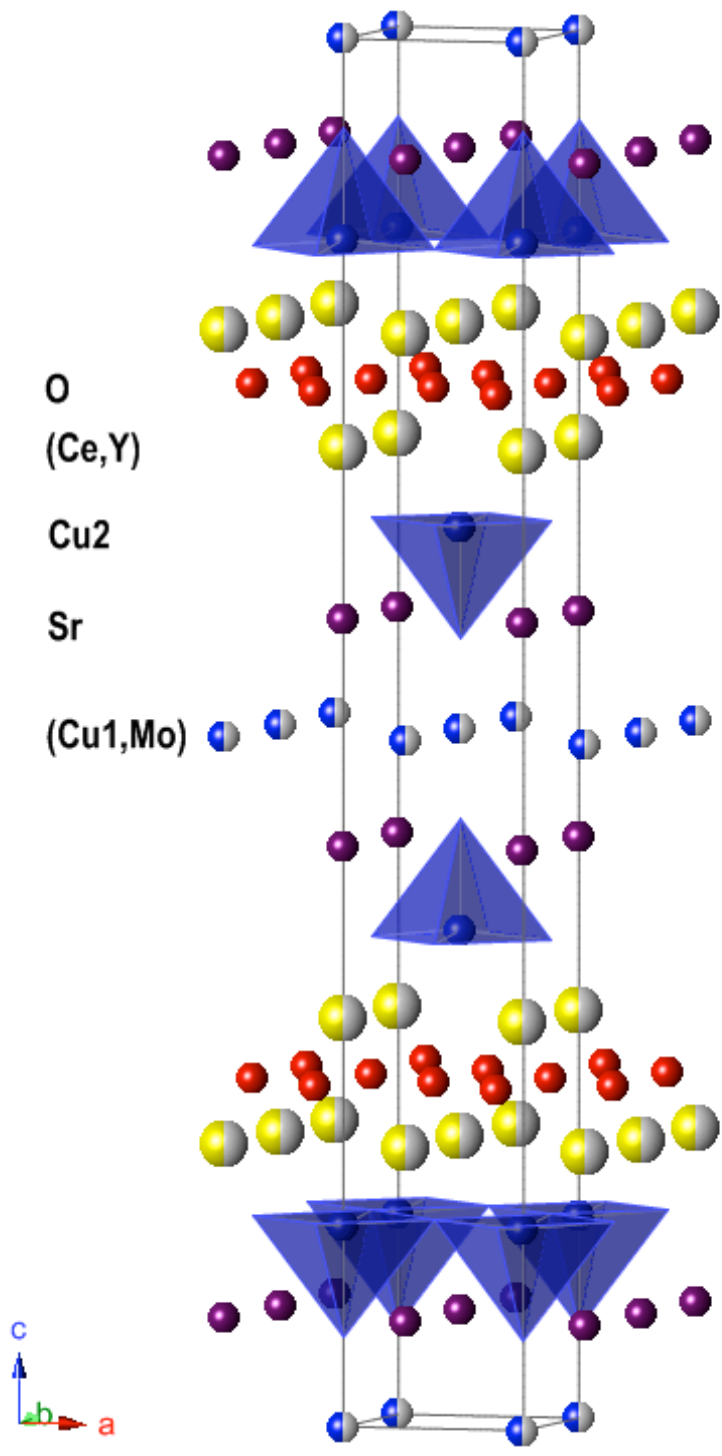

Figure 1

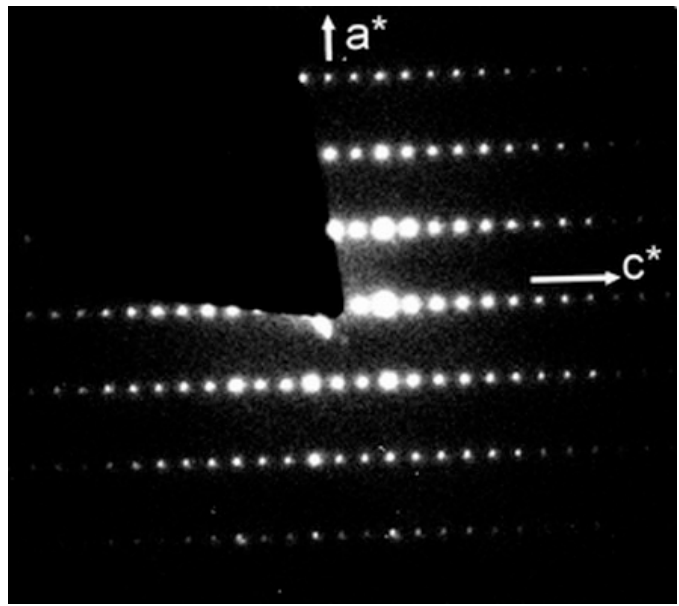

Figure 2a

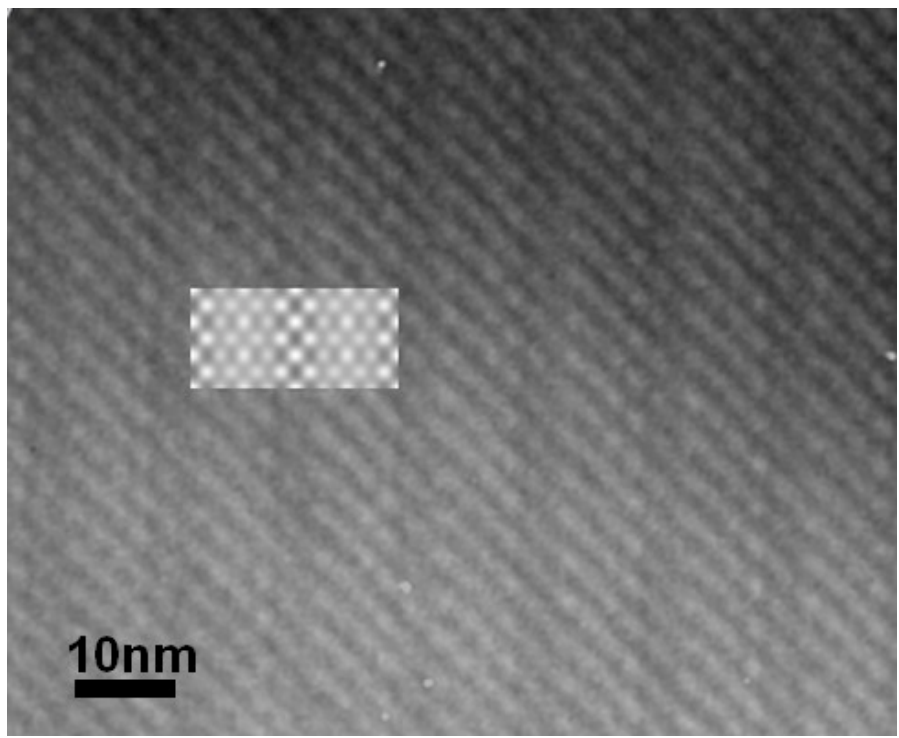

Figure 2b

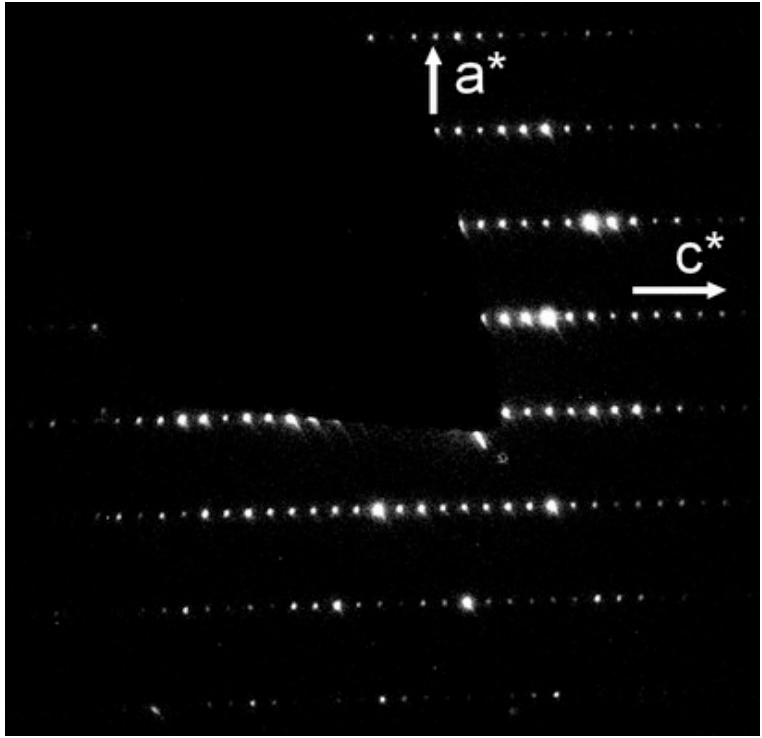

Figure 3a

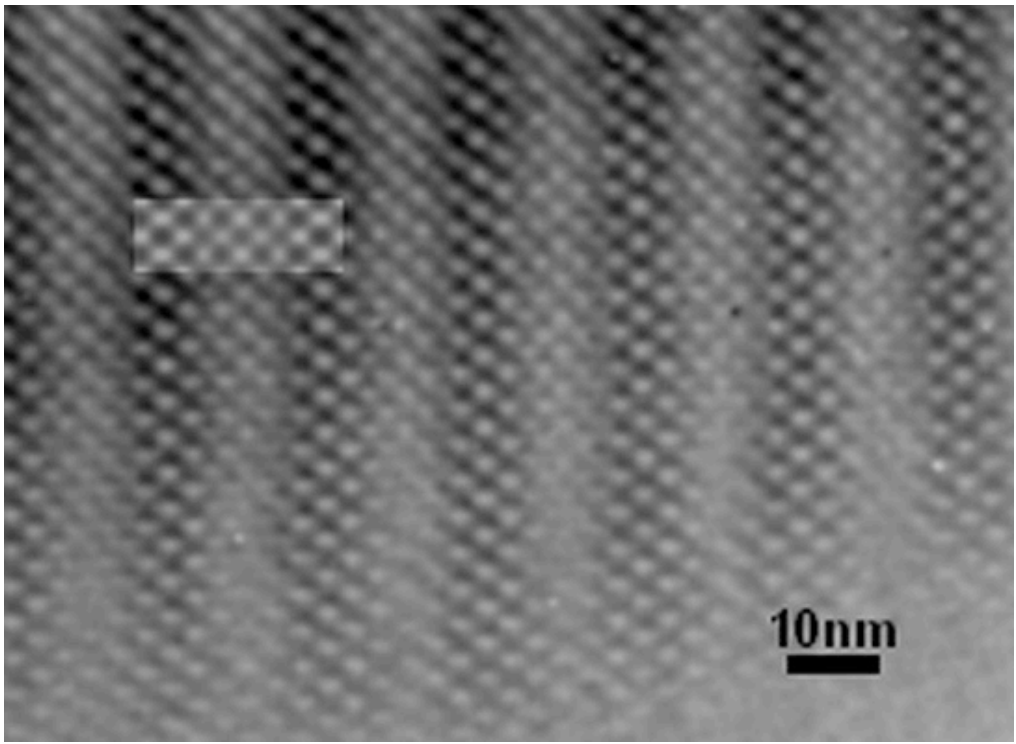

Figure 3b

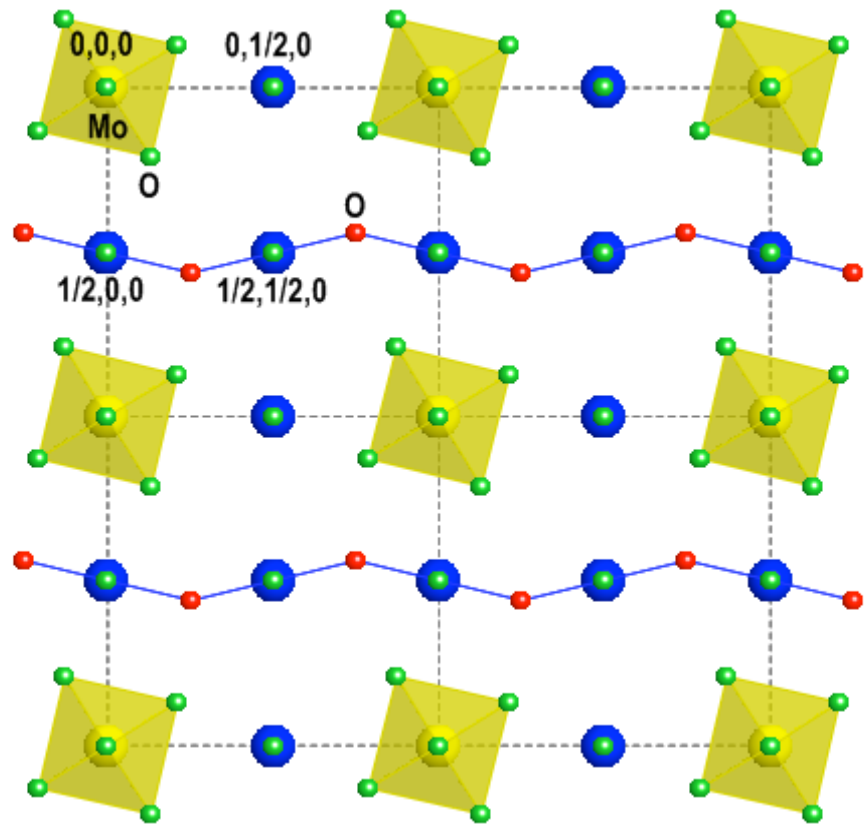

Figure 4

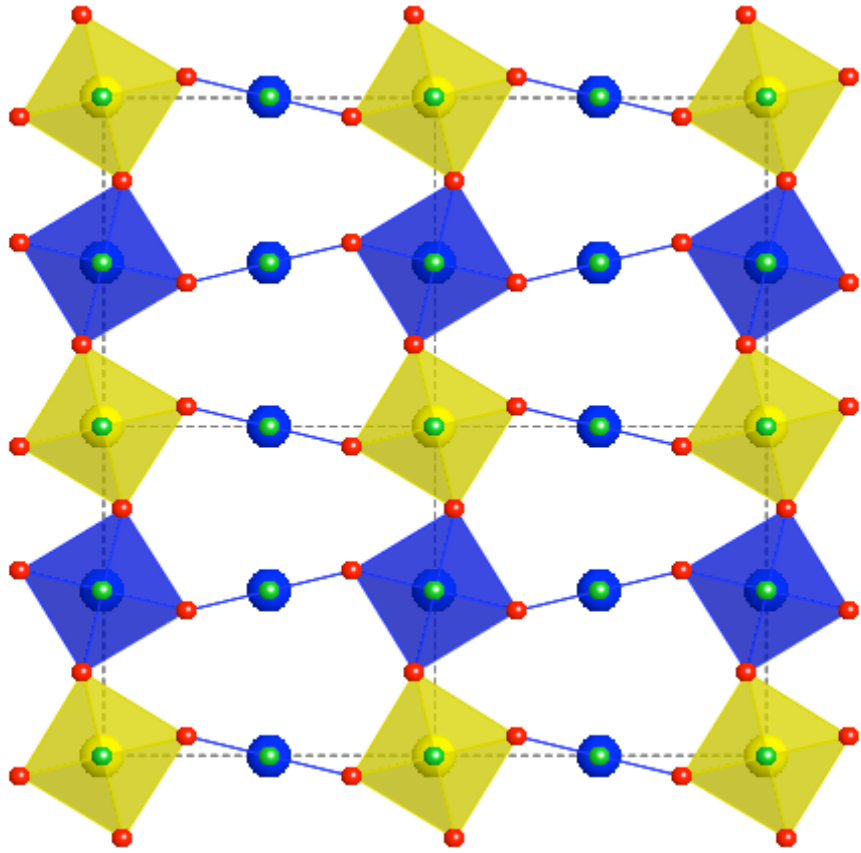

Figure 5